\documentclass{jetpl}
\onecolumn

\usepackage{graphics} 
\usepackage{epsfig}
\usepackage{multicol}
\usepackage{amsmath}
\begin{document}
\title{On Theory of the Phonon Perturbed Superradiance}

\rtitle{On Theory of the Phonon Perturbed Superradiance\ldots}

\sodtitle{On Theory of the Phonon Perturbed Superradiance}

\author{A.\,P.\,Saiko\/\thanks{saiko@ifttp.bas-net.by}}

\rauthor{A.\,P.\,Saiko}

\sodauthor{Saiko}

\address{Institute of Solid State Physics and Semiconductors, Belarus Academy of Sciences}

\abstract{The paper examines superradiance in impurity crystals in
the field of a coherent phonon wave excited by two ultrashort
laser pulses via Raman scattering processes at the moment of
preparation of the initial state of an ensemble of emitters. It is
shown that by varying the power of the excitation pulses and their
mutual direction of propagation, one can control the superradiance
parameters and extract data on the electron-phonon coupling
constant and its anisotropy.}

\maketitle
\section{INTRODUCTION}
The problem of spontaneous coherent emission (Dicke
superradiance~\cite{bib:1}) in impurity crystals taking into
account the interaction of emitters, impurity particles, and
thermal phonons, has been studied by a number of researchers
[2-7]. Weak emitter-phonon coupling generally leads to
temperature-dependent decrease in the superradiance intensity
~\cite{bib:5,bib:6}. But with strong adiabatic electron-phonon
coupling, for which the energy spectrum of the  system acquires a
set of electron vibrational levels consisting  of a zero phonon
line and vibrational repetitions, there is a finite probability of
emission of superradiance pulses not only in the zero-phonon
transition but also in the vibronic transitions, with the
intensity of these pulses, their delay times, and durations
strongly depending on the electron-phonon coupling parameter and
the temperature of the sample(whose dimensions are smaller than
the radiation wavelength)~\cite{bib:2,bib:7}. Similar results were
obtained in Refs. 3 and 4 using Thomson model ~\cite{bib:8}, which
describes Brillouin scattering of light in a resonant medium with
the participation of a single absorbed (or emitted) phonon.

Experimentally, superradiance on electron-vibrational transitions
of $O^{-}_{2}$ molecules was realized in the polar dielectrics
$KCL:O^{-}_{2}$ (see Refs. 9 and 10). Naboikin et al.
~\cite{bib:5} studied Dicke superradiance in mixed molecular
crystals of diphenyl with pyrene on pyrene centers, which
typically display weak electron-phonon coupling.

\section{STATEMENT OF THE PROBLEM}

Obviously, the parameters of superradiance can be effectively
controlled by exciting a coherent phonon wave, say by a
time-dependent Raman process, instead of exciting thermal phonons,
which because of their insignificant population are ineffective at
low temperatures. In this case the experimental setup changes
somewhat: to prepare the initial superradiance state at time $t =
0$ there must be, in addition to an ultrashort pulse of frequency
$\omega_{1}$, a second ultrashort pulse of frequency $\omega_{2}$
propagating at an angle to the first. Then, in addition to optical
excitation of impurity centers in the sample at frequencies
$\omega_{1}$ and $\omega_{2}$, a coherent phonon wave is excited
at the difference frequency $\omega_{1}-\omega_{2}=\Omega$, and
this wave modulates the frequency of the quantum  transitions of
the emitters (as in the case in which only thermal phonons
participate). Changing the direction of propagation of the
excitation laser pulses and tuning the difference frequency
$\omega_{1}-\omega_{2}$ to resonance with different phonon modes,
one can not only control the superradiance dynamics but also use
the behavior of the characteristic superradiance parameters to
determine the features of electron-phonon coupling of localized
electron states with the excited phonon modes and the anisotropy
of this coupling. Note that the role of a coherent phonon wave in
the light echo phenomenon was studied by Wilson et al.
 ~ \cite{bib:11} (see also Ref. 12).

We study an ensemble  of optical emitters\t impurity atoms or
molecules imbedded in a crystal lattice. A natural approach to
describing the electron states of impurity particles in resonant
phenomena is to use the idealized  scheme  of a two-level quantum
system ~\cite{bib:1}. We assume that each such two-level system
interacts with radiation field and the coherent phonon wave
(excited in this case by ultrashort laser pump radiation). The
interaction of the quantum levels and thermal phonons can be
ignored because the phonon modes of the lattice are "frozen"  if
low temperatures are used in the experimental setup. It is
legitimate to study a selectively excited coherent phonon wave by
classical means, i.e., by assuming that the wave has an amplitude
$q$, a frequency $\Omega$, and a phase $\phi$.

We write the Hamiltonian of the system of optical emitters
interacting with the radiation field and the coherent phonon wave
as
\begin{equation}
 H=H_{p}+H_{ph}+H_{int}+V(t),
 \label{eq:1}
\end{equation}
Where
$$
H_{p}=\omega_{0}\sum_{j}{S^{z}_{j}},\  \
$$
$$
H_{ph}=\sum_{{k}}{\omega_{k}a^{+}(k)a(k)},
$$

$$
H_{int}=\sum_{j,\{k\}}{[g_{j}(k)a(k)(S^{+}_{j}+S^{-}_{j})+H.c.]},
$$

$$
V(t)=\lambda\sum_{j}{S^{z}_{j}cos(\Omega t+\phi_{j})}.
 $$

Here $a(k)$ is the quantized amplitude of the radiation field for
the mode $(\textbf{k},s)$ with the frequency $\omega_{k}$ and
polarization $\textbf{e}_{s}$, $S^{\pm,z}_{j}$ is the pseudospin
variable, which obeys the commutation relations for angular
momentum and describes the $j$th ($j=1,2,...,N$) two level atom
with energy splitting $\omega_{0}$, and  $g_{j}(k)$ and $\lambda$
are the electron-photon and electron-phonon coupling constants:
$$g_{j}(k)=g(k)exp(i\textbf{k}\cdot \textbf{r}_{j})=-i\sqrt{2\pi c|\textbf{k}
|/V}(\textbf{d}\cdot\textbf{e}_{s})exp(i\textbf{k}\cdot
\textbf{r}_{j}),$$ With $\textbf{d}$ and $\textbf{r}_{j}$ the
dipole moment and radius vector of the $j$th particle and $V$ the
quantization volume of the radiation field, and $\phi_{j}$ the
phase of the phonon wave at the point occupied by the $j$th
particle. Throughout this paper we take $\hbar=1$.

\section{ DERIVATION OF THE MASTER EQUATION}

The statistical operator of a system described by the Hamiltonian
(\ref{eq:1}) satisfies the Liouville equation (the evolution of
the system is studied over time intervals shorter than the time of
irreversible dephasing of the polarization of the resonant medium)

\begin{equation}
 i\frac{d\rho}{dt}=[H,\rho],
 \label{eq:2}
\end{equation}
which the canonical transformation
\begin{equation}
 \rho\rightarrow\rho'+U^{+}\rho U,
 \label{eq:3}
\end{equation}
reduces to
\begin{equation}
 i\frac{d\rho'}{dt}=[H_{int}(t),\rho'],
 \label{eq:4}
\end{equation}
where
\begin{equation}
 U=exp\left\{i\int^{t}dt'[H_{p}+H_{ph}+V(t')]\right\},
 \label{eq:5}
\end{equation}
\begin{eqnarray}
 H_{int}=\sum_{j,\{k\}}{g_{j}(k)a(k)\left\{S^{+}_{j}exp\left[i\left(\omega_{0}t+\frac{\lambda}{\Omega}sin(\Omega
 t+\phi_{j})\right)\right]+H.c.\right\}exp[-i\omega(k)t]+H.c.}
 \label{eq:6}
\end{eqnarray}
To identify the contributions of the harmonics whose frequencies
are integral multiplies of $\Omega$, it is convenient to expand
the exponential function in (\ref{eq:6}) in a Bessel-function
series based on formula
$$
e^{iasinx}=\sum^{\infty}_{n=-\infty}{J_{n}{(a)}e^{inx}}.
 $$
Then
\begin{eqnarray}
 H_{int}=\sum_{j,\{k\},n}{g_{j}(k)J_{n}\left(\frac{\lambda}{\Omega}\right)a(k)\times [S^{+}_{j}exp[i(\omega_{0}-\omega(k)+n\Omega)t]\times exp(in\phi_{j})+H.c.]+H.c.}
 \label{eq:7}
\end{eqnarray}
Equation (\ref{eq:4}) can be transformed in a standard manner into
the integro-differential equation
\begin{equation}
 \frac{d\rho'}{dt}+i[H_{int}(t),\rho'(0)]=-\int^{t}_{0}[H_{int}(t),[H_{int}(t'),\rho'(t')]]dt'.
\label{eq:8}
\end{equation}
Finding the trace in the photon-field variables of both sides of
Eq. (8), we arrive at equations for the reduced density matrix
$\sigma=Tr_{ph}[\rho'(t)]$, which describes the evolution of the
quantum states of the impurity subsystem only:
\begin{equation}
 \frac{d\sigma}{dt}=-\int^{t}_{0}d\tau
 Tr_{ph}[H_{int}(t),[H_{int}(t-\tau),\rho'(t-\tau)]].
\label{eq:9}
\end{equation}
Here we have used the fact that $Tr_{ph}([H_{int}(t),\rho'(0)])=0$
which follows from the factorization of $\rho'$ at $t=0$: it is
assumed that initially the system is prepared in a state which
density matrix
\begin{equation}
\rho'(0)=\sigma(0)|0\rangle\langle0|, \label{eq:10}
\end{equation} where $|0\rangle\langle0|$ is the
density matrix of the radiation field (we may assume that for the
frequencies in the optical range the radiation field is in the
vacuum state $|0\rangle$ with zero temperature).

Since in the case of superradiance the field has the shortest
correlation time (proportional to the reciprocal of
optical-frequency bandwidth), in a certain sense the field can be
interpreted as a reservoir for the atomic system. The interaction
of this system with such a wide band reservoir rapidly wipes out
any memory of the system about its past, with the result that the
temporal behavior of the optical emitters is Markovian. This
justifies replacing $\rho'(t-\tau)$ by  $\rho'(t)$ in (\ref{eq:9})
and extending the upper limit of the $\tau$-interval to $\infty$.
Moreover, we assume that $\rho'(t)$ can be written as
$$
\rho'(t)=\sigma(t)|0\rangle\langle0|+\Delta\rho',
$$
where $\Delta\rho'$ is at least of order $H_{int}$. Then to second
order in $H_{int}$ we obtain a closed equation for the reduced
matrix density:
\begin{equation}
 \frac{d\sigma}{dt}=-\int^{t}_{0}d\tau
 Tr_{ph}[H_{int}(t),[H_{int}(t-\tau),\sigma(t)|0\rangle\langle0|]].
\label{eq:11}
\end{equation}
After substituting the expression (\ref{eq:7}) for $H_{int}$ into
the righthand side of Eq.(11), expanding commutators, taking the
trace, and integrating with respect $\tau$ we arrive at
$$
 \frac{d\sigma}{dt}=\sum_{j,l,\{k\},n,m}{J_{n}\left(\frac{\lambda}{\Omega}\right)J_{m}\left(\frac{\lambda}{\Omega}\right)}
  \times \left\{\frac{}{}exp[i(n-m)\Omega t]exp(in\phi_{j})exp(-im\phi_{l})\right.
\times[g_{j}^{*}(k)g_{l}(k)[S^{+}_{l},S^{-}_{j}\sigma]
 $$
$$
\times\left[\frac{iP}{\omega_{0}-\omega(k)+m\Omega}+\pi\delta[\omega_{0}-\omega(k)+m\Omega]\right]
 +g_{j}(k)g_{l}^{*}(k)[\sigma S^{-}_{l},S^{+}_{j}]
 $$
$$
\left.\times\left[\frac{iP}{\omega_{0}+\omega(k)+m\Omega}+\pi\delta[\omega_{0}+\omega(k)+m\Omega]\right]\right\}
 $$
\begin{equation}
+H.c.\equiv-\Lambda(t)\sigma.
 \label{eq:12}
\end{equation}
In the discussion that follows we ignore the principal part
(denoted by P), which yield only a small imaginary term, i.e.,
causes a small shift in the resonant frequency, and the
nonresonant terms containing the factor
$\delta[\omega_{0}+\omega(k)+m\Omega]$.

The characteristic time it takes the optical emitters to go into a
coherent state is proportional to $\|\Lambda\|^{-1}$ (here $\|...
\|$ stands for the operator's value in frequency units) and is
much longer than the period $T=2\pi/\Omega$ of the rapid harmonic
"jitter" superimposed on the less rapid evolution process of
spontaneous coherent emission. Hence it is natural to decompose
the real motion, described by the density matrix $\sigma$, into
slow motion averaged over the period $T$ and rapid "jitter". It is
convenient to realize the decomposition procedure by employing the
method of "temporal projection" operators~\cite{bib:13}:
\begin{equation}
\sigma(t)\rightarrow\tilde{\sigma}(t)=\frac{1}{T}\int_{0}^{T}dt\sigma(t)\equiv
P^{t}\sigma(t), \label{eq:13}
\end{equation}
where the projection operator $P^{t}$ performs the averaging of
rapidly varying quantities over the period $T$. Next we introduce
the operator $Q^{t}=1-P^{t}$. Then the decomposition of the
density matrix into the slowly and rapidly varying parts can be
written as
\begin{equation}
\sigma=P^{t}\sigma+Q^{t}\sigma. \label{eq:14}
\end{equation}

Successively applying the operators $P^{t}$and $Q^{t}$ to Eq.
(12), we arrive at two differential equations instead of one
~\cite{bib:13}:
\begin{equation}
\frac{P^{t}\sigma}{dt}=-P^{t}\Lambda(t)P^{t}\sigma-P^{t}\Lambda(t)Q^{t}\sigma
 \label{eq:15}
\end{equation}
\begin{equation}
\frac{Q^{t}\sigma}{dt}=-Q^{t}\Lambda(t)P^{t}\sigma-Q^{t}\Lambda(t)Q^{t}\sigma
 \label{eq:16}
\end{equation}
On the basis of Eq. (16) we can formally represent the solution
for the rapidly varying (oscillating) part of the density matrix
in the form
\begin{equation}
Q^{t}\sigma=-\int^{t}dt'Q^{t'}\Lambda(t')\tilde{\sigma}(t')-\int^{t}dt'Q^{t'}\Lambda(t')\sigma(t')
\label{eq:17}
\end{equation}
Using (17) in Eq. (15) for the slowly varying part of the density
matrix, we see that the order of smallness in $\Lambda$ of the
$P^{t}\Lambda Q^{t}\sigma$ is no less than two (or no less than
four in $H_{int}$), so that the term can be ignored. As a result,
averaging Eq. (12) over the "jitter" reduces, according to (15),
to averaging the operator $\Lambda(t)$ over the period
$2\pi/\Omega$:
\begin{equation}
 \frac{d\tilde{\sigma}}{dt}=\pi\sum_{j,l,n}{J_{n}^{2}\left(\frac{\lambda}{\Omega}\right)exp[in(\phi_{j}-\phi_{l})]}
 \times\sum_{\{k\}}{|g(k)|^{2}exp[i\textbf{k}\cdot(\textbf{r}_{j}-\textbf{r}_{l})]}
\times\delta(\omega_{0}-\omega(k)+n\Omega)[S_{j}^{-}\tilde{\sigma},S_{l}^{+}]+H.c.
 \label{eq:18}
\end{equation}

This expression constitutes the most general form of the master
equation for superradiance in the presence of a coherent phonon
wave. However, to use this equation that can be compared with
experimental data, we introduce a "coarsening" procedure. The
point here is that real samples containing emitters are much
larger that the radiation wavelength: $k_{0}R\gg1$ (here
$k_{0}=\omega/c$ and $R=max|\textbf{r}_{j}-\textbf{r}_{l}|$).
Hence $exp[i\textbf{k}\cdot(\textbf{r}_{j}-\textbf{r}_{l})]$ must
be averaged over the ensemble of particles. We apply a similar
procedure to the factor $exp[in(\phi_{j}-\phi_{l})]$, assuming
that the phase difference $\delta\phi_{jl}=\phi_{j}-\phi_{l}$ is
uniformly distributed over the interval $0$ to $2\pi$ with density
$1/2\pi$. The result of ensemble averaging,
$$\overline{exp[i\textbf{k}\cdot(\textbf{r}_{j}-\textbf{r}_{l})]}\equiv\Gamma(|\textbf{k}|),$$
Depends on the shape of the sample, while averaging over the phase
spread $\delta\phi_{jl}$ results in the exponential function
$exp[in\delta\phi_{jl}]$ being replaced by the Kronecker symbol:
\begin{equation}
\frac{1}{2\pi}\int_{0}^{2\pi}\delta\phi_{jl}exp(in\delta\phi_{jl})=\delta_{n,0}.
\label{eq:19}
\end{equation}
Allowing for (19) and passing in (18) from the sum over ${k}$ to
an integral with respect to the wave vector $k$ and a sum  over
the polarization s of the photon field,
$$\sum_{\{k\}}...\rightarrow\sum_{s}\frac{V}{(2\pi)^{3}}\int d^{3}k...,$$
we obtain
\begin{equation}
\frac{d\tilde{\sigma}}{dt}=-\gamma\Gamma(k_{0})J_{0}^{2}\left(\frac{\lambda}{\Omega}\right)\sum_{j,l}(S_{j}^{+}S_{l}^{-}\tilde{\sigma}-2S_{l}^{-}\tilde{\sigma}S_{j}^{+}+\tilde{\sigma}S_{j}^{+}S_{l}^{-}),
\label{eq:20}
\end{equation}
where $\gamma=2|\textbf{d}|^{2}\omega^{3}/3c^{3}$.

Thus the dependence of n-phonon processes (n=1,2,...) on the
randomly distributed phase shifts $\delta\phi_{jl}$ (with constant
density $\frac{1}{2\pi}$ must lead to averaging their contribution
to zero when superradiance is initiated in large samples. The
entire modulation effect of a coherent phonon wave on optical
centers in this case resides in the factor $J_{0}^{2}(\lambda /
\Omega)$, which renormalizes the electron photon coupling
constant. This factor is similar to the Debye-Waller factor, which
characterizes the reduction in the unshifted component (the
zero-phonon line) of an optical transition when elastic scattering
of thermal phonons is taken into account. However, in contrast to
the case of thermal phonons, a coherent phonon wave does not lead
to homogeneous dephasing of the polarization of the optical
centers.

\section{RESULTS AND DISCUSSION}

The standard form~\cite{bib:14}of the operator part of Eq. (20)
makes it possible to immediately write an (approximate) expression
for the intensity $I$ of superradiance under conditions in which
all emitters are initially completely inverted and the transverse
macroscopic polarization is zero, i.e.,
$Tr_{p}[S^{z}_{j}\tilde{\sigma}(0)]=\frac{1}{2}$ and
$Tr_{p}[S^{+}_{j}S^{-}_{l}\tilde{\sigma}(0)]=0$:
$$I(t)=-\omega_{0}\sum_{j}\frac{d}{dt}Tr_{p}[S^{z}_{j}\tilde{\sigma}(t)]=\frac{N\omega_{0}}{4\tau_{R}}sech^{2}\left(\frac{t-t_{D}}{2\tau_{R}}\right),$$
where the renormalized delay time $t_{D}$ and duration $\tau_{R}$
of the superradiance pulse are defined as
$$\tau_{R}=\tau_{R}^{(0)}J_{0}^{-2}\left(\frac{\lambda}{\Omega}\right),$$
$$t_{D}=\tau_{R}lnN=t_{D}^{(0)}J_{0}^{-2}\left(\frac{\lambda}{\Omega}\right)$$
(here $\tau_{R}^{(0)}=[\gamma\Gamma(k_{0})N]^{-1}$) and
$t_{D}^{(0)}$ are the values of the parameters in the absence of a
coherent phonon wave).

The renormalization factor $J_{0}^{2}(\lambda / \Omega)$, which
enters  into the expression for $t_{D}$, $\tau_{D}$, and $I$
because of the modulation of the quantum states of the emitters by
a coherent phonon wave, leads (since $J_{0}^{2}(\lambda /
\Omega)<1$ for $\lambda / \Omega\neq0$) to an increase  in the
time of formation of a superradiance pulse, the spreading of the
pulse, and a decrease in the peak intensity. The advantage of
using a coherent phonon wave is that not only does it insure
effective control over  the superradiance process but it also
allows extracting information about the strength of the coupling
of the electron states of the emitters and the various phonon
modes thanks to the possibility  of selectively exciting these
modes (and simultaneous "freezing" of thermal phonons caused by
low temperatures). These facts are illustrated by Fig. 1, which
depicts the time dependence of the intensity $I$ for the different
values of the total energy $W$ of the excitation pulses.
\begin{figure}[h]
\centering\includegraphics*[width=8cm,height=8cm,angle=0]{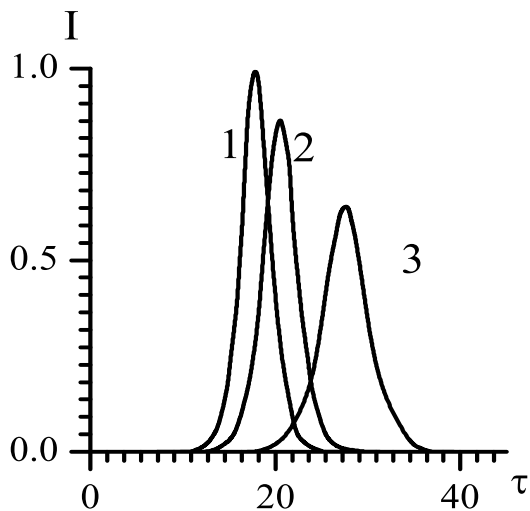}
\caption{The intensity $I$ of Dicke superradiance (normalized to
$N\omega_{0}/4\tau_{R}^{0}$) as a function of
$\tau=t/\tau^{(0)}_{R}$ for three values of the energy $W$ of the
excitation laser pulses: 1), $W=0$; 2), $W=1.25\times10^{-3}J$;
and 3), $W=2.25\times10^{-3}J$ ($N=10^{18}$).}
\end{figure}
Here we have allowed for the fact that the dimensionless parameter
$\lambda/\Omega$ can be written as ~\cite{bib:11}
$\lambda/\Omega=q_{0}\sqrt{2Sm\Omega/\hbar}$, where $q_{0}=AW$ is
the phonon amplitude, $S$ is a constant (for the $a_{1g}$-mode
naphthalene, $\Omega=1385$\,cm$^{-1}$, $S=0.01$,
$A=0.3\times10^{-13}$dyn$^{-1}$, and $m\simeq10^{3}m_{p}$, with
$m_{p}$ the proton mass). Figure 2 illustrates the dependence of a
superradiance pulse on the electron-phonon coupling constant
$\lambda / \Omega$ for the fixed value of $W$).
\begin{figure}[h]
\centering\includegraphics*[width=8cm,height=8cm,angle=0]{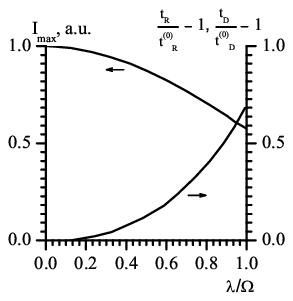}
\caption{The peak intensity $I_{max}$, the delay time $t_{D}$, and
the superradiance pulse duration $\tau_{R}$ as a functions of the
electron-phonon coupling constant $\lambda/\Omega$.}

\end{figure}
Obviously, in the latter case it is possible to study the
anisotropy of the electron-phonon coupling by varying the
direction of propagation of the phonon wave (i.e, by changing the
direction in which the excitation laser pulses are applied).

\vfill\eject

\end{document}